\documentclass[12pt]{article}

\usepackage{graphicx}
\usepackage{amstext}

\def\ltwid{\mathrel{\raise.3ex\hbox{$<$\kern-.75em\lower1ex\hbox{$\sim$}}}}
\def\gtwid{\mathrel{\raise.3ex\hbox{$>$\kern-.75em\lower1ex\hbox{$\sim$}}}}

\def\square{\kern1pt\vbox{\hrule height 1.2pt\hbox{\vrule width 1.2pt\hskip 3pt
   \vbox{\vskip 6pt}\hskip 3pt\vrule width 0.6pt}\hrule height 0.6pt}\kern1pt}
\def\overleftrightarrow#1{\vbox{\ialign{##\crcr
     $\leftrightarrow$\crcr\noalign{\kern-1pt\nointerlineskip}
     $\hfil\displaystyle{#1}\hfil$\crcr}}}

\begin{document}

\begin{titlepage}

\begin{flushright}
CCTP-2012-09 \\ UFIFT-QG-12-05
\end{flushright}

\vskip 1cm

\begin{center}
{\bf Graviton Propagator in a General Invariant Gauge on de Sitter}
\end{center}

\vskip 1cm

\begin{center}
P. J. Mora$^{1*}$, N. C. Tsamis$^{2\dagger}$ and R. P.
Woodard$^{1,\ddagger}$
\end{center}

\vskip 1cm

\begin{center}
\it{$^{1}$ Department of Physics, University of Florida \\
Gainesville, FL 32611, UNITED STATES}
\end{center}

\begin{center}
\it{$^{2}$ Institute of Theoretical Physics \& Computational Physics \\
Department of Physics University of Crete \\
GR-710 03 Heraklion, HELLAS}
\end{center}

\vskip 1cm

\begin{center}
ABSTRACT
\end{center}
We construct the graviton propagator on de Sitter background
in the one parameter family of exact, de Sitter invariant gauges. 
Our result takes the form of a universal spin two part and a 
gauge dependent spin zero part. Scalar equations are derived for
the structure functions of each part. There is no de Sitter
invariant solution for either structure function, although the
de Sitter breaking contribution to the spin zero part may drop
out for certain choices of the gauge parameter. Our results 
imply that de Sitter breaking is universal for the graviton
propagator, and hence that there is an error in the contrary
results derived by analytic continuation of average gauge
fixing techniques.

\begin{flushleft}
\end{flushleft}

\begin{flushleft}
$^*$ e-mail: pmora@phys.ufl.edu \\
$^{\dagger}$ e-mail: tsamis@physics.uoc.gr \\
$^{\ddagger}$ e-mail: woodard@phys.ufl.edu
\end{flushleft}

\end{titlepage}

\section{Introduction}

The idea of gauge invariance has been central to theories of vector
and tensor fields. The fact that physical results should be gauge
independent allows one to choose the best option for a gauge in 
terms of simplicity and convenience. However, this freedom cannot 
be exploited unless the propagator is known in a variety of different 
gauges, and that information is only available for a handful of
backgrounds. For gravitons on flat space background there are no
subtleties to imposing a general Poincar\'e invariant gauge, and
one loop results are known for the graviton self-energy \cite{Capper}
and for the vacuum polarization \cite{LW}. In contrast, all graviton
loop computations on de Sitter background \cite{TW1,TW2,TW3,MW1,KW1}
have been performed in a single gauge \cite{TW4,RPW}.

The reason for this is two subtleties in gauge fixing on de Sitter
background that became apparent from problems in early work on the
subject \cite{AM,Folacci}:
\begin{itemize}
\item{There is a topological obstacle to adding de Sitter invariant
gauge fixing terms \cite{MTW1}; and}
\item{Analytic continuation techniques can fail when infrared 
divergences are present \cite{MTW2}.}
\end{itemize}
The first problem invalidates most of the solutions that have been
obtained by adding de Sitter invariant gauge fixing terms \cite{INVPROP};
the math was right but there was a subtle physics error with gauge 
fixing. It should still be possible to enforce exact gauge conditions 
which are de Sitter invariant, but the second problem means that one 
cannot trust exact gauges which are alleged to be enforced by taking 
singular limits of average gauges \cite{FH}.

The more reliable procedure for exact gauge conditions is to construct
the propagator directly. This has recently been accomplished 
\cite{MTW3,KMW} for de Donder gauge,
\begin{equation}
D^{\nu} h_{\mu\nu} - \frac12 D_{\mu} h^{\nu}_{~\nu} = 0 \; ,
\end{equation}
where $h_{\mu\nu}$ is the graviton field, whose indices are raised and 
lowered by the de Sitter background metric, and $D_{\mu}$ denotes the
covariant derivative operator in the de Sitter background. The purpose
of this paper is to extend this result to a general de Sitter invariant
gauge,
\begin{equation}
\label{gc}
D^{\nu} h_{\mu\nu} - \frac{\beta}{2} D_\mu h^{\nu}_{~\nu} = 0 \; .
\end{equation} 

We follow the analysis of the $\beta = 1$ case \cite{MTW3} by splitting
the propagator into a spin zero part (which depends upon $\beta$)
and a spin two part (which is independent of $\beta$). Both of these 
will be written in terms of projection operators, which automatically 
enforce the gauge condition, acting on structure functions. We then write
down a graviton propagator equation which obeys the gauge condition on both
index groups, with the appropriate projection operator standing on the
right hand side. It is then simple to derive scalar differential equations
for the two structure functions, which can be solved using previous
techniques.

In the next section we present our notation and give a summary of 
the formalism of working on de Sitter space. In section 3 we review 
previous work on general scalar and vector propagators and explain 
the method of integrating propagators. The main result is derived in 
Section 4 where we present the projection operators and structure 
functions for both the spin zero and spin two parts. Section 5 
compares our results with those obtained by the suspect procedure of
taking singular limits of gauge fixing terms \cite{FH}.


\section{The de Sitter Background}

In this section we will set the notation for this paper by defining 
the graviton field and presenting the formalism for working on a de 
Sitter background. The graviton propagator we will obtain corresponds 
to a gravitational field described by the Lagrangian:
\begin{equation}
\mathcal{L}_G \equiv \frac1{16\pi G} \Bigl( R - (D\!-\!2)
\Lambda\Bigr) \sqrt{-g} \; .
\end{equation}
Here $D$ is the dimension of spacetime, $\Lambda$ is the cosmological 
constant, which is related to the Hubble constant by $\Lambda=(D-1)H^2$ 
and $G$ is Newton's constant. We define the graviton field 
$h_{\mu\nu}(x)$ by subtracting the de Sitter background metric 
$\overline{g}_{\mu\nu}(x)$ from the full metric $g_{\mu\nu}(x)$:
\begin{equation}
g_{\mu\nu}(x) = \overline{g}_{\mu\nu}(x) + \kappa h_{\mu\nu}(x).
\end{equation}
Here $\kappa^2 \equiv 16 \pi G$ is the loop counting parameter 
of quantum gravity. Because all tensorial constructions will 
henceforth involve the background metric, we drop the bar and 
employ $g_{\mu\nu}(x)$ to represent the de Sitter background. 
Graviton indices are raised and lowered with this background field, 
$h^\mu_\sigma=g^{\mu\rho}h_{\rho\sigma}$ and the covariant derivative 
operator $D_\mu$ is constructed with respect to this background field. 
This operator serves to define the Lichnerowicz operator which will 
be useful in Section \ref{sec grav prop}:
\begin{eqnarray}
\label{lich}
{\bf D}^{\mu\nu\rho\sigma}&\equiv& D^{(\rho}g^{\sigma)(\mu}D^{\nu)}-\frac{1}{2}\Big[g^{\mu\nu}D^\rho D^\sigma+g^{\rho\sigma}D^\mu D^\nu\Big]\nonumber\\
&+&\frac{1}{2}\Big[g^{\mu\nu}g^{\rho\sigma}-g^{\mu(\rho}g^{\sigma)\nu}\Big]\square+(D-1)\Big[\frac{1}{2}g^{\mu\nu}g^{\rho\sigma}-g^{\mu(\rho}g^{\sigma)\nu}\Big]H^2,
\end{eqnarray}
where we have used the de Sitter result for the Riemann tensor 
$R_{\alpha\beta\mu\nu} = H^2 ( g_{\alpha\mu} g_{\beta\nu} - g_{\beta\mu}
g_{\alpha\nu})$. From now on we will employ the notation that 
par\-en\-the\-siz\-ed indices are symmetrized and $\square\equiv 
g^{\mu\nu}D_\mu D_\nu$ is the covariant d'Alembertian operator.

Different physicists have different preferences about what is most appropriate choice for a de Sitter space manifold. Many prefer to work on the full de Sitter manifold because of the high level symmetry of de Sitter space. In this work we make no ${\it a\;priori}$ assumptions about de Sitter symmetry, we do however wish to, from the perspective of inflationary cosmology, preserve the symmetries of isotropy and homogeneity, or the $(E3)$ vacuum, for a spatially flat space. Hence we prefer to do our calculations in the open conformal submanifold of de Sitter, also known as the ``cosmological patch''. Here the ranges for the coordinates $x^\mu=(x^0,x^i)$ are:
\begin{equation}
-\infty<x^0<0\;\;\;\;\;\;\text{and}\;\;\;\;\;\;-\infty<x^i<\infty\;\;\;\;\;\;\text{for}\;\;\;i=1,\dots,(D-1)\;.
\end{equation}
In this coordinates the metric, being conformal to that of flat space, takes a simple form :
\begin{equation}
ds^2\equiv g_{\mu\nu}dx^\mu dx^\nu = a^2\Big[-(dx^0)^2+d\vec{x}\cdot d\vec{x}\Big]=a^2\eta_{\mu\nu}dx^\mu dx^\nu \;,
\end{equation}
where $a=-1/Hx^0$ is the scale factor and $\eta_{\mu\nu}$ is the Lorentz metric.

Our calculation will generally involve de Sitter invariant and non-invariant quantities. Manifest infrared effects produce divergences that will break de Sitter invariance. However much of this work will contain invariant quantities, and when present it is useful to express them in terms of the de Sitter invariant length function $y(x,x')$,
\begin{equation}
y(x;x') \equiv a a' H^2 \Biggl[\Bigl\Vert \vec{x} \!-\! \vec{x}\;'
\Vert^2 - \Bigl( \vert x^0\!-\! x'^0 \vert \!-\! i \varepsilon
\Bigr)^2 \Biggr] \; . \label{ydef}
\end{equation}
Except for the factor of $i \varepsilon$ (whose purpose is to
enforce Feynman boundary conditions) the de Sitter length function
can be expressed as follow in terms of the geodesic length
$\ell(x;x')$ from $x^{\mu}$ to ${x'}^{\mu}$,
\begin{equation}
y(x;x') = 4 \sin^2\Bigl( \frac12 H \ell(x;x')\Bigr) \; .
\end{equation}
Once we have this de Sitter invariant function we can construct a 
basis of de Sitter invariant bi-tensors. Since $y(x;x')$ is de 
Sitter invariant so too are covariant derivatives of it. This basis 
consists of the first three derivatives of $y(x;x')$ along with the 
metrics $g_{\mu\nu}(x)$ and $g_{\mu\nu}(x')$ \cite{KW2}:
\begin{eqnarray}\label{yb1}
\frac{\partial y(x;x')}{\partial x^{\mu}} & = & H a \Bigl(y
\delta^0_{\mu}
\!+\! 2 a' H \Delta x_{\mu} \Bigr) \; , \label{dydx} \\
\frac{\partial y(x;x')}{\partial {x'}^{\nu}} & = & H a' \Bigl(y
\delta^0_{\nu}
\!-\! 2 a H \Delta x_{\nu} \Bigr) \; , \label{dydz} \\
\label{yb3}
\frac{\partial^2 y(x;x')}{\partial x^{\mu} \partial {x'}^{\nu}} & = &
H^2 a a' \Bigl(y \delta^0_{\mu} \delta^0_{\nu} \!+\! 2 a' H
\Delta x_{\mu} \delta^0_{\nu} \!-\! 2 a \delta^0_{\mu} H \Delta
x_{\nu} \!-\! 2 \eta_{\mu\nu}\Bigr) \; . \qquad \label{dydxdx'}
\end{eqnarray}
Here and subsequently we define $\Delta x_{\mu} \equiv \eta_{\mu\nu}
(x \!-\!x')^{\nu}$. Acting covariant derivatives generates more basis
tensors, for example \cite{KW2},
\begin{equation}
\frac{D^2 y(x;x')}{Dx^{\mu} Dx^{\nu}}
= H^2 (2 \!-\!y) g_{\mu\nu}(x) \quad , \quad \label{covdiv}
\frac{D^2 y(x;x')}{D {x'}^{\mu} D {x'}^{\nu}} = H^2 (2 \!-\!y)
g_{\mu\nu}(x') \; .
\end{equation}
The contraction of any pair of the basis tensors also produces more
basis tensors \cite{KW2},
\begin{eqnarray}
g^{\mu\nu}(x) \frac{\partial y}{\partial x^{\mu}} \frac{\partial
y}{\partial x^ {\nu}} & = & H^2 \Bigl(4 y - y^2\Bigr) =
g^{\mu\nu}(x') \frac{\partial y}{
\partial {x'}^{\mu}} \frac{\partial y}{\partial {x'}^{\nu}} \; ,
\label{contraction1}\\
g^{\mu\nu}(x) \frac{\partial y}{\partial x^{\nu}} \frac{\partial^2 y}{
\partial x^{\mu} \partial {x'}^{\sigma}} & = & H^2 (2-y) \frac{\partial y}{
\partial {x'}^{\sigma}} \; ,
\label{contraction2}\\
g^{\rho\sigma}(x') \frac{\partial y}{\partial {x'}^{\sigma}}
\frac{\partial^2 y}{\partial x^{\mu} \partial {x'}^{\rho}} & = & H^2
(2-y) \frac{\partial y}{\partial x^{\mu}} \; ,
\label{contraction3}\\
g^{\mu\nu}(x) \frac{\partial^2 y}{\partial x^{\mu} \partial
{x'}^{\rho}} \frac{\partial^2 y}{\partial x^{\nu} \partial {x'}^{\sigma}}
& = & 4 H^4 g_{\rho\sigma}(x') - H^2 \frac{\partial y}{\partial
{x'}^{\rho}} \frac{\partial y}{\partial {x'}^{\sigma}} \; ,
\label{contraction4}\\
g^{\rho\sigma}(x') \frac{\partial^2 y}{\partial x^{\mu}\partial
{x'}^{\rho}} \frac{\partial^2 y}{\partial x^{\nu} \partial {x'}^{\sigma}}
& = & 4 H^4 g_{\mu\nu}(x) - H^2 \frac{\partial y}{\partial x^{\mu}}
\frac{\partial y}{\partial x^{\nu}} \; . \label{contraction5}
\end{eqnarray}


\section{General Scalar and Vector Propagators}

It is our goal to find an expression for the graviton propagator 
in terms of covariant derivative projection operators acting on 
some scalar functions. The method we will use to obtain these 
scalar functions consists of first finding a differential equation 
they must obey and then integrating these scalar propagator equations. 
Additionally we will also require the use of vector propagators to 
facilitate our calculation when imposing our gauge condition on the 
graviton propagator equation. The purpose of this section is to 
present the general solution for the propagators of a scalar and 
vector field with masses $M_S$ and $M_V$ respectively and explain 
the method of integrating propagators \cite{MTW1,MTW2}.

\subsection{Scalar Propagators}

Let us define $b_A \equiv (D-1)/2$, and consider a general scalar 
field with mass-squared $M_S^2 = (b_A^2 - b^2) H^2$ described by 
the Lagrangian,
\begin{equation}
{\cal L}_S = -\frac{1}{2} \partial_\mu \varphi \partial_\nu \varphi 
g^{\mu\nu} \sqrt{-g} - \frac{1}{2} M_{S}^2 \varphi^2 \sqrt{-g} \; .
\end{equation}
The associated propagator $i\Delta_b(x;x')$ obeys the equation,
\begin{equation}
\label{scaleqn}
\Big[ \square \!+\! b^2 H^2 \!-\! b_A^2 H^2 \Big] i\Delta_b(x;x')
= \frac{i\delta^D(x \!-\!x')}{\sqrt{-g}} \; .
\end{equation}
The propagator will be de Sitter invariant for a positive mass-squared 
which corresponds to $b < b_A$. Its expansion for $b = \nu$ is:
\begin{eqnarray}
\label{Sinv}
\lefteqn{i \Delta_\nu^{\text{dS}}(x;x') = \frac{H^{D-2}}{(4\pi)^{D/2}}
\Biggr\{ \Gamma\Big(\frac{D}{2} \!-\!1\Big) \Big(\frac{4}{y}
\Big)^{\frac{D}{2}-1} \!-\! \frac{\Gamma(\frac{D}{2}) \Gamma(1 \!-\!
\frac{D}{2})}{\Gamma(\frac{1}{2} \!+\! \nu)\Gamma(\frac{1}{2} \!-\! \nu)}
\times\Big. \sum_{n=0}^{\infty} } \nonumber \\
& & \hspace{-.7cm} \times \Biggr.\Biggr[ \frac{\Gamma(\frac{3}{2} \!+\! \nu 
\!+\! n) \Gamma(\frac{3}{2} \!-\! \nu \!+\! n)}{\Gamma(3 \!-\! \frac{D}{2} 
\!+\!n) (n \!+\! 1)!} \Big(\frac{y}{4}\Big)^{n-\frac{D}{2}+2} 
\!\!\!-\! \frac{\Gamma(b_a \!+\! \nu \!+\! n)\Gamma(b_A \!-\! \nu \!+\! n)}{
\Gamma(\frac{D}{2} \!+\! n) n!} \Big(\frac{y}{4}\Big)^{n} \Biggr] \Biggr\} .
\qquad 
\end{eqnarray}
For the case $b \geq b_A$ the naive mode sum becomes infrared divergent 
and requires a de Sitter breaking infrared correction \cite{AF,MTW3}. 
This correction term is added in such a way to preserve the symmetries 
of homogeneity and isotropy mentioned before. The expansion of the de 
Sitter breaking part is,
\begin{eqnarray}
\label{Snoninv}
\lefteqn{\Delta_\nu^{\text{IR}}(x;x')=\frac{H^{D-2}}{(4\pi)^{D/2}}\frac{\Gamma\Big(\nu\Big)\Gamma(2\nu)}{\Gamma(b_A)\Gamma(\nu+\frac{1}{2})}\times\theta(\nu-b_A)}\nonumber\\
&&\times\displaystyle\sum^{\nu-b_A}_{N=0}\frac{(aa')^{\nu-b_A-N}}{\nu-b_A-N}\displaystyle\sum^N_{n=0}\Bigg(\frac{a}{a'}+\frac{a'}{a}\Bigg)^n\displaystyle\sum^{\left(\frac{N-n}{2}\right)}_{m=0}C_{Nnm}(y-2)^{N-n-2m} \;,
\end{eqnarray}
where the coefficients $C_{Nnm}$ are,
\begin{eqnarray}
\label{Coef}
\lefteqn{C_{Nnm}=\frac{(-\frac{1}{4})^N}{m!n!(N-n-2m)!}\times\frac{\Gamma(b_A+N+n-\nu)}{\Gamma(b_A+N-\nu)}}\nonumber\\
&&\times\frac{\Gamma(b_A)}{\Gamma(b_A+N-2m)}\times\frac{\Gamma(1-\nu)}{\Gamma(1-\nu+n+2m)}\times\frac{\Gamma(1-\nu)}{\Gamma(1-\nu+m)}\;.
\end{eqnarray}
The full propagator is defined as the limit as $\nu$ approaches $b$ 
of the sum of expressions (\ref{Sinv}) and (\ref{Snoninv}):
\begin{eqnarray}
\label{scalarProp}
i\Delta_b(x;x') = \displaystyle\lim_{\nu\rightarrow b}
\Bigg[i\Delta_\nu^{\rm dS}(x;x') + \Delta_\nu^{\text{IR}}(x;x')\Bigg] \; .
\end{eqnarray}

We close this subsection by showing a technique we will require later for the frequent case of integrating scalar propagator equations when the source term contains another scalar propagator. Suppose that a ``propagator'' $i\Delta_{bc}(x;x')$ obeys the equation
\begin{equation}
\label{bceqn}
\Big[\square+(b^2-b_A^2)H^2\Big]i\Delta_{bc}(x;x')=i\Delta_c(x;x').
\end{equation}
The solution to this equation is \cite{MTW1,MTW2}:
\begin{equation}
\label{bcprop}
i\Delta_{bc}(x;x')=\frac{1}{(b^2-c^2)H^2}\Big[i\Delta_c(x;x')-i\Delta_b(x;x')\Big]=i\Delta_{cb}(x;x')\;.
\end{equation}
For the special case $b=c$ the solution involves a derivative,
\begin{equation}
\label{bbprop}
i\Delta_{bb}(x;x')=-\frac{1}{2bH^2}\frac{\partial}{\partial b}i\Delta_b(x;x').
\end{equation}
We can apply the same procedure when the source is an integrated propagator,
\begin{equation}
\label{intprop}
\Big[\square+(b^2-b_A^2)H^2\Big]i\Delta_{bcd}(x;x')=i\Delta_{cd}(x;x').
\end{equation}
The solution to this equation, which is manifestly symmetric under the interchange of any two indices, is given by:
\begin{eqnarray}
\label{bcdprop}
\lefteqn{i\Delta_{bcd}(x;x')=\frac{i\Delta_{bd}(x;x')-i\Delta_{bc}(x;x')}{(c^2-d^2)H^2}}\\
& & \hspace{0.15cm}=\frac{(d^2-c^2)i\Delta_b(x;x')+(b^2-d^2)i\Delta_c(x;x')+(c^2-b^2)i\Delta_d(x;x')}{(b^2-c^2)(c^2-d^2)(d^2-b^2)H^4}\;.
\end{eqnarray}
Again, when two indices agree the solution involves a derivative,
\begin{equation}
\label{2int}
i\Delta_{bcc}(x;x')=-\frac{1}{2cH^2}\frac{\partial}{\partial c}i\Delta_{bc}(x;x')=\frac{i\Delta_{cc}(x;x')-i\Delta_{bc}(x;x')}{(b^2-c^2)H^2}.
\end{equation}
If all three indices are equal then we obtain,
\begin{eqnarray}
\label{bbb}
i\Delta_{bbb}(x;x')&=&-\frac{1}{2bH^2}\frac{\partial}{\partial b}i\Delta_{bc}(x;x')\Bigg\vert_{c=b}\nonumber\\
&=&-\frac{1}{8b^3H^4}\Biggr[\frac{\partial}{\partial b}i\Delta_{b}(x;x')-b\left(\frac{\partial}{\partial b}\right)^2i\Delta_{b}(x;x')\Biggr].
\end{eqnarray}

\subsection{Vector Propagators}

In this subsection we will present the results for a general vector propagator corresponding to a vector field with mass-squared $M_V^2$ described by the following Lagrangian
\begin{equation}
{\cal L}_V=-\frac{1}{2}\partial_\mu A_\rho\partial_\nu A_\sigma g^{\mu\nu}g^{\rho\sigma}\sqrt{-g}-\frac{1}{2}\Bigg[(D-1)H^2+M^2_V\Bigg]A_\rho A_\sigma g^{\rho\sigma}\sqrt{-g}\;.
\end{equation}
>From this equation we see that its associated propagator $i\Bigr[\mbox{}_{\mu} \Delta_{\rho}\Bigr](x;x')$ obeys the equation
\begin{equation}
\label{veceqn}
\Big[\square - (D-1)H^2 - M^2_V\Big]i\Bigr[\mbox{}_{\mu} \Delta_{\rho}\Bigr](x;x') =\frac{g_{\mu\rho}i\delta^D(x-x')}{\sqrt{-g}}\;.
\end{equation}

It is important to note that at this point this is the full vector propagator. However it can be decomposed, without making any assumptions about de Sitter invariance, into a longitudinal part and a transverse part:
\begin{equation}
\label{vectorprop}
i\Bigr[\mbox{}_{\mu} \Delta_{\rho}\Bigr](x;x')=i\Bigr[\mbox{}_{\mu} \Delta^L_{\rho}\Bigr](x;x')+i\Bigr[\mbox{}_{\mu} \Delta^T_{\rho}\Bigr](x;x')\;.
\end{equation}

Both terms in the above expression were obtained as projector operators constructed in terms of covariant derivatives acting on scalar structure functions.
The longitudinal part can be simply written as,
\begin{equation}
\label{vectorL}
i\Bigr[\mbox{}_{\mu} \Delta^L_{\rho}\Bigr](x;x')=\frac{\partial}{\partial x^\mu}\frac{\partial}{\partial x'^\rho}\Big[{\cal S}_L(x;x')\Big]\;.
\end{equation}
The transverse part, although a little more complicated, can be written as,
\begin{equation}
\label{vectorT}
i\Bigr[\mbox{}_{\mu} \Delta^T_{\rho}\Bigr](x;x')={\cal P}^{\alpha\beta}_{\mu}(x)\times{\cal P}^{\gamma\delta}_{\rho}(x')\times{\cal Q}_{\alpha\gamma}(x;x')\times\Big[{\cal R}_{\beta\delta}(x;x'){\cal S}_T(x;x')\Big]\;,
\end{equation}
where
\begin{equation}
{\cal R}_{\beta\delta}(x;x')\equiv-\frac{1}{2H^2}\frac{\partial^2 y(x;x')}{\partial x^\beta \partial x'^\delta}\;,
\end{equation}
and the operators are defined by
\begin{eqnarray}
{\cal P}^{\alpha\beta}_\mu(x)&\equiv&\delta^\beta_\mu D^\alpha-\delta^\alpha_\mu D^\beta\;,\nonumber\\
{\cal Q}_{\alpha\gamma}(x;x')&\equiv&-\frac{1}{2H^2}\frac{D}{D x^\alpha}\frac{D}{D x'^\gamma}\;.
\end{eqnarray}
Note that this choice corresponds to enforcing transversality when acting ${\cal P}^{\alpha\beta}_\mu(x){\cal P}^{\rho\sigma}_\nu(x')$ on any 4-index symmetric bi-tensor function of $x$ and $x'$. Having found the tensorial structure of both parts we now only need the solution for the structure functions ${\cal S_L}$ and ${\cal S_T}$.

We can find an equation for the longitudinal structure function by taking the divergence of the full propagator equation (\ref{veceqn}) and using equations (\ref{vectorprop})-(\ref{vectorT}). The result is
\begin{equation}
\label{Leqn}
\Big[\square-M^2_V\Big]\square'{\cal S}_L(x;x')=-\frac{i\delta^D(x-x')}{\sqrt{-g}}\;.
\end{equation}
This equation can be easily solved for $\square'{\cal S}_L$ by comparing it with (\ref{scaleqn}),
\begin{equation}
\square'{\cal S}_L(x;x')=-i\Delta_b(x;x')\;\;,\;\;\;\;\;\text{for}\;\;\;\;\;\;\;b^2=\left(\frac{D-1}{2}\right)^2-\frac{M^2_V}{H^2}.
\end{equation}
One more integration using relations (\ref{bceqn})-(\ref{bcprop}) yields the desired result,
\begin{equation}
\label{solSL}
{\cal S}_L(x;x')=\frac{1}{M^2_V}\Big[i\Delta_A(x;x')-i\Delta_b(x;x')\Big]=-i\Delta_{Ab}(x;x')\;.
\end{equation}

Having obtained the solution for the longitudinal part we can now 
derive an equation for the transverse part by substituting this 
solution in the full propagator equation (\ref{veceqn}) \cite{TW5},
\begin{eqnarray}
\lefteqn{\Big[\square - (D-1)H^2\ - M^2_V\Big]i\Bigr[\mbox{}_{\mu} 
\Delta^T_{\rho}\Bigr](x;x') =}\nonumber\\
& & \hspace{4cm} \frac{g_{\mu\rho}i\delta^D(x-x')}{\sqrt{-g}}+\frac{\partial}{\partial x^\mu}\frac{\partial}{\partial x'^\rho}i\Delta_A(x;x')\;.
\end{eqnarray}
From this it can be shown \cite{MTW3} that the transverse structure 
function obeys the equation,
\begin{equation}
\Big[\square-(D-2)H^2-M^2_V\Big]{\cal S}_T(x;x')=-2H^2i\Delta_{BB}(x;x')\;.
\end{equation}
Applying relations (\ref{bceqn})-(\ref{bcprop}) we conclude that the transverse structure function is,
\begin{equation}
\label{solST}
{\cal S}_T(x;x')=\frac{2H^2}{M^2_V}i\Delta_{BB}(x;x')+\frac{2H^2}{M^4_V}\Big[i\Delta_B(x;x')-i\Delta_c(x;x')\Big]\;,
\end{equation}
where $c=\sqrt{\left(\frac{D-3}{2}\right)^2-\frac{M^2_V}{H^2}}$ and the $B-$type propagator corresponds to the value $b_B=\left(\frac{D-3}{2}\right)$.


\section{The Graviton Propagator}\label{sec grav prop}

This section comprises our main result. We will use the techniques 
presented in the previous section, namely that for integrating 
propagator equations and the method of projection operators to 
calculate the graviton propagator in the one parameter family of 
exact, de Sitter invariant gauges. We consider the most general 
invariant extension (\ref{gc}) of de Donder gauge. As with the 
vector propagator we will decompose the graviton propagator into 
two parts: a spin zero and a spin two part. Similarly we will write 
these contributions as projection operators acting on two scalar 
structure functions ${\cal S}_0(x;x')$ and ${\cal S}_2(x;x')$, respectively.

\subsection{Imposing a General Invariant Gauge}

When the exact gauge condition (\ref{gc}) is imposed, the propagator must obey the same condition on either coordinate and the corresponding index group:
\begin{eqnarray}
\label{propg}
\Bigg[\delta^\mu_\lambda D^\nu-\frac{\beta}{2}D_\lambda g^{\mu\nu}(x)\Bigg]\times i\Bigr[\mbox{}_{\mu\nu} \Delta_{\rho\sigma}\Bigr](x;x')&=&0\;,\\
\Bigg[\delta^\rho_\lambda D'^\sigma-\frac{\beta}{2}D'_\lambda g^{\rho\sigma}(x')\Bigg]\times i\Bigr[\mbox{}_{\mu\nu} \Delta_{\rho\sigma}\Bigr](x;x')&=&0\;.
\end{eqnarray}
It is these two equations that will determine the projection operators we seek in order to decompose the full graviton propagator into its spin 0 and spin 2 parts,
\begin{equation}
\label{spins}
i\Bigr[\mbox{}_{\mu\nu} \Delta_{\rho\sigma}\Bigr](x;x')=i\Bigr[\mbox{}_{\mu\nu} \Delta^0_{\rho\sigma}\Bigr](x;x')+i\Bigr[\mbox{}_{\mu\nu} \Delta^2_{\rho\sigma}\Bigr](x;x').
\end{equation}
So far this is completely analogous to the case of the vector propagator 
where the conditions enforced were longitudinality and transversality.

The spin zero part of the propagator can be written in terms of an operator which is a linear combination of longitudinal and trace terms,
\begin{equation}
\label{0prop}
i\Bigr[\mbox{}_{\mu\nu} \Delta^0_{\rho\sigma}\Bigr](x;x')={\cal P}_{\mu\nu}(x)\times {\cal P}_{\rho\sigma}(x')\Big[{\cal S}_0(x;x')\Big].
\end{equation}
To determine this projector operator we consider the following linear combination of operators,
\begin{equation}
\label{rawP}
{\cal P}_{\mu\nu}(x)=D_\mu D_\nu+c_1g_{\mu\nu}\square+c_2g_{\mu\nu}H^2.
\end{equation}
The constants $c_1$ and $c_2$ are determined by enforcing the exact gauge condition on the propagators. Substituting equations (\ref{0prop}) and (\ref{rawP}) into (\ref{propg}) we obtain $c_1=(2-\beta)/(D\beta-2)$ and $c_2=2(D-1)/(D\beta-2)$. Thus our spin zero projector takes the form,
\begin{eqnarray}
\label{newP}
\lefteqn{{\cal P}_{\mu\nu}(x) = D_\mu D_\nu \!+\! g_{\mu\nu} 
\left(\frac{2 \!-\! \beta}{D \beta \!-\! 2}\right) 
\Biggl[\square \!+\! 2 \left(\frac{D \!-\! 1}{2 \!-\! \beta} \right)H^2 
\Biggr] \; , } \\
& & = D_{\mu} D_{\nu} \!+\! \frac{g_{\mu\nu}}{D \!-\! 2} \Bigl[ \square
\!+\! 2 (D \!-\! 1) H^2\Bigr] \!+\! g_{\mu\nu} \frac{2 (D \!-\! 1) 
(1 \!-\! \beta)}{(D \!-\! 2) (D \beta \!-\! 2)} \Bigl[ \square \!+\!
D H^2 \Bigr] \; . \label{2ndform} \qquad
\end{eqnarray}

Our analysis for the spin two part of the propagator is facilitated 
by the fact that this part is not affected by the gauge fixing 
parameter. When $\beta=1$ we recover exact de Donder gauge and the 
graviton propagator in this gauge has been published in \cite{MTW3}. 
The result is,
\begin{equation}
\label{2prop}
i\Bigr[\mbox{}_{\mu\nu} \Delta^2_{\rho\sigma}\Bigr](x;x')=\frac{1}{4H^4}{{\bf P}_{\mu\nu}}^{\alpha\beta}(x)\times {{\bf P}_{\rho\sigma}}^{\kappa\lambda}(x')\Big[{\cal R}_{\alpha\kappa}(x;x'){\cal R}_{\beta\lambda}(x;x'){\cal S}_2(x;x')\Big]\;,
\end{equation}
where the projection operator is defined by
\begin{eqnarray}
\label{2proj}
\lefteqn{{{\bf P}_{\mu\nu}}^{\alpha\beta}(x)=\frac{1}{2}\left(\frac{D-3}{D-2}\right)\Big\{-\delta^\alpha_{(\mu}\delta^\beta_{\nu)}\Big[\square-DH^2\Big]\Big[\square-2H^2\Big]}\Big.\nonumber\\
& & \hspace{1cm} + \Big. g_{\mu\nu}g^{\alpha\beta}\Big[ \frac{\square^2}{D-1}-H^2\square+2H^4 \Big] -  \frac{D_{(\mu}D_{\nu)}}{D-1}\Big[\square+2(D-1)H^2\Big]g^{\alpha\beta}\Big.\nonumber\\
& & \hspace{1cm} + \Big. 2D_{(\mu}\Big[\square+H^2\Big]\delta^{(\alpha}_{\nu)}D^{\beta)} - \left(\frac{D-2}{D-1}\right)D_{(\mu}D_{\nu)}D^{(\alpha}D^{\beta)}\Big.\nonumber\\
& & \hspace{1cm} - \Big. \frac{g_{\mu\nu}}{D-1}\Big[\square+2(D-1)H^2\Big]D^{(\alpha}D^{\beta)}\Big\}\;.
\end{eqnarray}
Now that we have the form of the graviton propagator in equations (\ref{0prop})and (\ref{2prop}), we only need to determine the corresponding scalar structure functions. In order to obtain equations for them we need to derive an equation for the propagator. To accomplish this we act the Lichnerowicz operator defined in (\ref{lich}) and impose the general gauge condition (\ref{propg}),
\begin{eqnarray}
\label{1propeqn}
\lefteqn{-D^{\mu\nu\rho\sigma}h_{\rho\sigma}=\frac{1}{2}(1-\beta)D^\mu D^\nu h^\sigma_\sigma+\frac{1}{2}\Big[\square-2H^2\Big]h^{\mu\nu}}\nonumber\\
& & \hspace{3.5cm} - \frac{(2-\beta)}{4}g^{\mu\nu}\Big[\square+\frac{2(D-3)}{(2-\beta)}H^2\Big]h^\sigma_\sigma\;.
\end{eqnarray}
From this expression we infer that the graviton propagator obeys the 
following equation,
\begin{eqnarray}
\label{rawpropeqn}
\lefteqn{\frac{(1-\beta)}{2}D_\mu D_\nu i\Bigr[\mbox{}_{\alpha}^\alpha \Delta_{\rho\sigma}\Bigr](x;x') +\frac{1}{2}\Big[\square-2H^2\Big]i\Bigr[\mbox{}_{\mu\nu} \Delta_{\rho\sigma}\Bigr](x;x')}\nonumber\\
& & \hspace{1.5cm} - \frac{(2-\beta)}{4}g_{\mu\nu}\Big[\square+\frac{2(D-3)}{(2-\beta)}H^2\Big]i\Bigr[\mbox{}_{\alpha}^\alpha \Delta_{\rho\sigma}\Bigr](x;x')=\text{R.H.S.}
\end{eqnarray}
Here the R.H.S., yet undetermined, includes terms that enforce the gauge condition on this side of the equation. One important point is that the left hand side obeys the gauge condition on $x'$ but not on $x$ due to the presence of the covariant operators. Hence the right hand side cannot be symmetric under the interchange of $x\leftrightarrow x'$ with the corresponding interchange of index groups. It can easily be seen that the term that spoils this symmetry is proportional to the trace of the propagator, so we achieve a more symmetric expression if we add a trace term to the left hand side of the equation to cancel the problematic term, namely,
\begin{eqnarray}
\label{newLHS}
\lefteqn{\frac{(1-\beta)}{2}D_\mu D_\nu i\Bigr[\mbox{}_{\alpha}^\alpha \Delta_{\rho\sigma}\Bigr](x;x') +\frac{1}{2}\Big[\square-2H^2\Big]i\Bigr[\mbox{}_{\mu\nu} \Delta_{\rho\sigma}\Bigr](x;x')}\nonumber\\
& & \hspace{1cm} - \frac{(2-\beta)}{4}g_{\mu\nu}\Big[\square+\frac{2(D-3)}{(2-\beta)}H^2\Big]i\Bigr[\mbox{}_{\alpha}^\alpha \Delta_{\rho\sigma}\Bigr](x;x')\nonumber\\
& & \hspace{2cm} + \frac{\beta}{4}\frac{(2-\beta)(D-2)}{(D\beta-2)}g_{\mu\nu}\Big[\square+\frac{2(D-1)}{2-\beta}H^2\Big]i\Bigr[\mbox{}_{\alpha}^\alpha \Delta_{\rho\sigma}\Bigr](x;x')\nonumber\\
\lefteqn{=\frac{(1-\beta)}{2}D_\mu D_\nu i\Bigr[\mbox{}_{\alpha}^\alpha \Delta_{\rho\sigma}\Bigr](x;x') + \frac{1}{2}\Big[\square-2H^2\Big]i\Bigr[\mbox{}_{\mu\nu} \Delta_{\rho\sigma}\Bigr](x;x')}\nonumber\\
&& + \frac{(2-\beta)}{2(D\beta-2)}g_{\mu\nu}\Big[(1-\beta)\square+\frac{2(D-3+\beta)}{(2-\beta)}H^2\Big]i\Bigr[\mbox{}_{\alpha}^\alpha \Delta_{\rho\sigma}\Bigr](x;x').
\end{eqnarray}
Now one can check that this equation obeys the general gauge condition on both $x$ and $x'$ making the right hand side symmetric under interchange of $x$ and $x'$. We will now derive the right hand side of this equation such that it also obeys the gauge condition (\ref{gc}). For this purpose we consider the following ansatz based on general tensor analysis:
\begin{eqnarray}
\label{ansatz}
\lefteqn{{\rm R.H.S.} = g_{\mu(\rho} g_{\sigma)\nu} 
\frac{i\delta^D(x \!-\! x')}{\sqrt{-g}} \!+\! c_1 g_{\mu\nu} g_{\rho\sigma}
\frac{i\delta^D(x \!-\! x')}{\sqrt{-g}} } \nonumber \\
& & \hspace{-.2cm} + \frac{c_2}{2} {D_\mu D'_\rho i\Bigr[\mbox{}_{\nu}
\Delta^T_{\sigma}\Bigr](x;x') \!+\! D_\nu D'_\rho i\Bigr[\mbox{}_{\mu}
\Delta^T_{\sigma}\Bigr] \atopwithdelims \{\} \!+\! D_\mu D'_\sigma 
i\Bigr[\mbox{}_{\nu} \Delta^T_{\rho}\Bigr](x;x') \!+\! D_\nu D'_\sigma 
i\Bigr[\mbox{}_{\mu}\Delta^T_{\rho}\Bigr]} \!+\! c_3 D_\mu D_\nu 
D'_\rho D'_\sigma S_L(x;x') \; . \qquad
\end{eqnarray}
Here we have used the transverse vector propagator $i[\mbox{}_{\nu}
\Delta^T_{\sigma}](x;x')$ which obeys the equation,
\begin{equation}
\label{Tveceqn}
\Big[\square + (D-1)H^2\Big]i\Bigr[\mbox{}_{\mu} \Delta^T_{\sigma}\Bigr](x;x') =\frac{g_{\mu\sigma}i\delta^D(x-x')}{\sqrt{-g}}+D_\mu D'_\sigma i\Delta_A(x;x')\;,
\end{equation}
The last term in this equation serves to enforce transversality which is immediate after we recall the equation the $A$-type propagator:
\begin{equation}\label{Aeqn}
b_A = \left(\frac{D-1}{2}\right) \qquad \Longrightarrow \qquad
\square i\Delta_A(x;x') = \frac{i\delta^D(x \!-\! x')}{\sqrt{-g}} \; .
\end{equation}
Similarly, the scalar function the four derivatives act upon in 
(\ref{ansatz}) must obey the relation,
\begin{equation}
\label{newScaleqn}
\Big[\square + \frac{2(D-1)}{2-\beta}H^2\Big]S_L(x;x') =-i\Delta_A(x;x')\;.
\end{equation}
We can integrate this equation as usual with the help of (\ref{bceqn}) to obtain,
\begin{equation}
\label{newScal}
S_L(x;x') = -\frac{(2-\beta)}{2(D-1)H^2}\Big[i\Delta_A(x;x')-i\Delta_N(x;x')\Big]\;,
\end{equation}
where the new $N$-type propagator obeys,
\begin{equation}
\label{NTypeEqn}
\Big[\square + \frac{2(D-1)}{2-\beta}H^2\Big]i\Delta_N(x;x') =\frac{i\delta^D(x-x')}{\sqrt{-g}}\;,
\end{equation}
with a corresponding value of $b^2_N=\frac{1}{4}(D-1)(D-1
+ \frac{8}{2-\beta})$. With the above relations, our ansatz 
(\ref{ansatz}) obeys the gauge condition provided we set 
$c_1 = \frac{-\beta}{(D\beta-2)}$, $c_2 = 1$ and $c_3 = 
\frac{2}{(2-\beta)}$. Hence our final form for the graviton 
propagator equation is,
\begin{eqnarray}
\label{newgravpropeqn}
\lefteqn{\frac{(1 \!-\! \beta)}{2} D_\mu D_\nu i\Bigr[
\mbox{}_{~\alpha}^\alpha \Delta_{\rho\sigma}\Bigr](x;x') \!+\! 
\frac{1}{2} \Big[\square \!-\!2 H^2 \Big] \, i\Bigr[\mbox{}_{\mu\nu} 
\Delta_{\rho\sigma}\Bigr](x;x') } \nonumber \\
& & \hspace{1.5cm} + g_{\mu\nu} \frac{(2 \!-\! \beta)}{2 (D \beta \!-\!2)}
\Big[(1 \!-\! \beta) \square \!+\! \frac{2 (D \!-\! 3 \!+\! \beta)}{(2 \!-\!
\beta)} H^2 \Big] \, i\Bigr[\mbox{}_{~\alpha}^\alpha 
\Delta_{\rho\sigma}\Bigr](x;x') \nonumber \\
& & \hspace{-.5cm} = g_{\mu(\rho} g_{\sigma)\nu} \frac{i\delta^D(x \!-\! 
x')}{\sqrt{-g}} \!-\! \frac{\beta g_{\mu\nu} g_{\rho\sigma} }{(D\beta \!-\! 
2)} \frac{i\delta^D(x \!-\! x')}{\sqrt{-g}} \nonumber \\
& & \hspace{-.2cm} + \frac{1}{2}{D_\mu D'_\rho i\Bigr[\mbox{}_{\nu}
\Delta^T_{\sigma}\Bigr](x;x') \!+\! D_\nu D'_\rho i\Bigr[\mbox{}_{\mu}
\Delta^T_{\sigma}\Bigr] \atopwithdelims \{\} \!+\! D_\mu D'_\sigma 
i\Bigr[\mbox{}_{\nu}\Delta^T_{\rho}\Bigr](x;x') \!+\! D_\nu D'_\sigma 
i\Bigr[\mbox{}_{\mu}\Delta^T_{\rho}\Bigr]} \!+\! \frac{2 D{\mu} D_{\nu} 
D'_{\rho} D'_{\sigma}}{(2 \!-\! \beta)} \, S_L(x;x') \; . \qquad
\end{eqnarray}

\subsection{The Structure Functions}

Our final task is to obtain the differential equations the two structure functions ${\cal S}_0(x;x')$ and ${\cal S}_2(x;x')$ obey and solve them using the technique presented in Section 3.

To find the equation for ${\cal S}_0(x;x')$ we just take the trace of the left hand side of our graviton propagator equation (\ref{newgravpropeqn}):
\begin{eqnarray}
\label{Trlhs}
\lefteqn{\left(\frac{D-2}{D\beta-2}\right)\Big[\frac{1}{2}(2-\beta)\square+(D-1)H^2\Big]i\Bigr[\mbox{}_{\alpha}^\alpha \Delta_{\rho\sigma}\Bigr](x;x')}\nonumber\\
& & \hspace{0.5cm}=\frac{(2-\beta)(D-2)(D-1)}{(D\beta-2)^2}\nonumber\\
& & \hspace{2cm} \times\Biggr[ \square+2\left(\frac{D-1}{2-\beta}\right)H^2 \Biggr]\Big[\square+DH^2\Big]{\cal P}^\prime_{\rho\sigma}{\cal S}_0(x;x')\;.
\end{eqnarray}
To arrive at this form we have used the full propagator (\ref{spins}) (the spin two part drops out because it is traceless) and substituted equations (\ref{0prop})-(\ref{rawP}). On the other hand, tracing the right hand side of (\ref{newgravpropeqn}) gives,
\begin{eqnarray}
\label{Trrhs}
g_{\rho\sigma}\Big[1-\frac{D\beta}{D\beta-2}\Big]\frac{i\delta^D(x-x')}{\sqrt{-g}}+0+\frac2{2-\beta}D'_\rho D'_\sigma\square S_L(x;x')
\;.
\end{eqnarray}
At this point we recall the expression we derived for the scalar function $S_L(x;x')$ in terms of the $A$-type and $N$-type propagators, (\ref{newScal}). It is easy to see, using equations (\ref{Aeqn}) and (\ref{NTypeEqn}), that acting the d'Alembertian operator on it gives $\square S_L(x;x')=-i\Delta_N(x;x')$. Hence (\ref{Trrhs}) becomes,
\begin{eqnarray}
\label{Trrhs2}
\lefteqn{-\frac{2}{(D\beta-2)}g_{\rho\sigma}(x')\frac{i\delta^D(x-x')}{\sqrt{-g}}-\frac{2}{2-\beta}D'_\rho D'_\sigma i\Delta_N(x;x')}\nonumber\\
& & \hspace{1.5cm} = -\frac{2}{(D\beta-2)}g_{\rho\sigma}(x')\Big[\square'+2\left(\frac{D-1}{2-\beta}\right)H^2\Big]i\Delta_N(x;x')\nonumber\\
& & \hspace{2cm}-\frac{2}{2-\beta}D'_\rho D'_\sigma i\Delta_N(x;x')\nonumber\\
& & \hspace{1.5cm} = -\frac{2}{2-\beta}\left\{ D'_\rho D'_\sigma+\left(\frac{2-\beta}{D\beta-2}\right)g_{\rho\sigma}(x')\right.\nonumber\\
& & \hspace{4cm}  \times\left.\Big[ \square'+2\left(\frac{D-1}{2-\beta}\right)H^2 \Big] \right\}i\Delta_N(x;x')\nonumber\\
& & \hspace{1.5cm} = -\frac{2}{2-\beta}{\cal P}'_{\rho\sigma}i\Delta_N(x;x').
\end{eqnarray}
Equating the traces of both sides of the propagator equation, (\ref{Trlhs}) and (\ref{Trrhs2}), gives us an equation for ${\cal S}_0$,
\begin{eqnarray}
\label{prem}
\frac{(2-\beta)(D-2)(D-1)}{(D\beta-2)^2}\Biggr[\square+2\left(\frac{D-1}{2-\beta}\right)H^2\Biggr]\Big[\square+DH^2\Big]{\cal S}_0(x;x')\nonumber\\
= -\frac2{2-\beta}i\Delta_N(x;x').
\end{eqnarray}
This equation can be simplified if we rewrite the source term as a delta function by acting the operator that inverts $i\Delta_N(x;x')$,
\begin{eqnarray}
\label{0eqn}
\lefteqn{\Big[\square+DH^2\Big]\Biggr[\square+2\left(\frac{D-1}{2-\beta}\right)H^2\Biggr]^2{\cal S}_0(x;x')}\nonumber\\
& & \hspace{2cm}= \frac{-2(D\beta-2)^2}{(2-\beta)^2(D-2)(D-1)}\frac{i\delta^D(x-x')}{\sqrt{-g}}\;.
\end{eqnarray}
This is the desired result for the equation for the spin zero structure function. Solving this equation finally gives the solution for the structure function ${\cal S}_0$,
\begin{equation}
\label{s0}
{\cal S}_0(x;x')=-\frac{2(D\beta-2)^2}{(2-\beta)^2(D-2)(D-1)}i\Delta_{WNN}(x;x')\;.
\end{equation}

From the definition of the doubly integrated propagator (\ref{2int}) 
we note that ${\cal S}_0$ involves the $W$-type propagator, 
$i\Delta_W(x;x')$, which corresponds to a tachyonic mass squared of 
$M_S^2=-DH^2$, and the $N$-type propagator, $i\Delta_N(x;x')$, which 
corresponds to a mass squared of $M_S^2=-2(\frac{D-1}{2-\beta})H^2$. 
As a consequence $i\Delta_W(x;x')$ will involve both a de Sitter 
invariant part and a gauge independent de Sitter breaking part. The 
propagator $i\Delta_N(x;x')$ has a gauge dependent mass-squared and 
breaks de Sitter invariance for $\beta<2$. This is in agreement with 
a previous result \cite{KMW} where explicit results were obtained for 
the de Sitter breaking part of the graviton propagator with $\beta=1$.

As we have already mentioned for the spin two sector of the propagator, 
this part is not affected by the gauge parameter. Thus we can impose 
again an exact de Donder gauge by setting $\beta=1$. The advantage of 
doing so is that the structure function ${\cal S}_2(x;x')$ has also 
been calculated in this gauge \cite{MTW3}. There it was shown that 
${\cal S}_2(x;x')$ obeys the equation,
\begin{eqnarray}
\label{2eqn}
\frac{1}{2}\square{\cal S}_2(x;x')=\left(\frac{4}{D-3}\right)^2\Big[i\Delta_{AA}(x;x')-2i\Delta_{AB}(x;x')+i\Delta_{BB}(x;x')\Big].
\end{eqnarray}
The solution follows from (\ref{intprop})-(\ref{bbb}),
\begin{eqnarray}
\label{s2}
{\cal S}_2(x;x')=\frac{32}{(D-3)^2}\Big[i\Delta_{AAA}(x;x')-2i\Delta_{AAB}(x;x')+i\Delta_{ABB}(x;x')\Big].
\end{eqnarray}
We remind the reader that the $B-$type propagator $i\Delta_B(x;x')$ has mass-squared $M^2_S=(D-2)H^2$ and thus is de Sitter invariant.


\section{Discussion}

We have constructed the graviton propagator on de Sitter background 
in the general family of exact de Sitter invariant gauges (\ref{gc}) 
parameterized by the constant $\beta$. Just as for the de Donder gauge 
case ($\beta = 1$), our result (\ref{spins}) takes the form of a 
transverse-traceless, spin two part (\ref{2prop}) plus a spin zero 
part (\ref{0prop}). Each part consists of a de Sitter invariant 
projector, which enforces the gauge condition, acting on a structure 
function. Neither the spin two projector (\ref{2proj}), nor the spin 
two structure function (\ref{s2}) is at all changed from the de Donder 
gauge solutions \cite{MTW3}. For general $\beta$, the spin zero 
projector is (\ref{newP}) and the spin zero structure function is
(\ref{s2}).

Scalar propagators break de Sitter invariance for any $M_S^2 \leq 0$
\cite{AF,MTW2}. Both structure functions break de Sitter invariance for 
all values of $\beta$. The spin two structure function (\ref{s2}) 
involves differences of scalar propagators for $M_S^2 = 0$ and $M_S^2 = 
(D -2 ) H^2$, and it has been shown that its de Sitter breaking
persists even after the spin two projectors (\ref{2proj}) have been
acted \cite{KMW}. {\it Hence the phenomenon of de Sitter breaking is 
universal.} The spin zero structure function (\ref{s0}) involves 
differences of the scalar propagators for $M_S^2 = -D H^2$ and 
$M_S^2 = -2(D-1)/(2 - \beta) H^2$. For the case of $\beta = 1$ it
has been shown that acting the spin zero projectors annihilates the
universal de Sitter breaking from the $M_S^2 = -DH^2$ propagator 
\cite{KMW}, and relation (\ref{2ndform}) reveals that this remains 
true for all $\beta$. So the spin zero part of the propagator is de
Sitter invariant for $\beta > 2$.

It should be noted that de Sitter breaking occurs as well in the
propagator which was constructed by adding a noninvariant gauge
fixing term \cite{TW4,RPW} that does not suffer from the topological
obstruction to invariant terms. Even though this gauge breaks de 
Sitter invariance, adding the compensating gauge transformation 
reveals a physical breaking of de Sitter symmetry \cite{Kleppe}. 
Indeed, the de Sitter breaking of this old propagator is precisely
the same in the transverse-traceless sector as what we have just
found for general $\beta$ \cite{KMW}. Our new propagator also gives
the same result for the linearized Weyl-Weyl correlator \cite{PMW}, 
because the correlators have been shown to agree for $\beta = 1$ 
\cite{PMTW}, and because the $\beta$-dependent, spin zero part drops 
out of the Weyl-Weyl correlator \cite{PMTW}.

Our universally de Sitter breaking results contrast sharply with the
universally de Sitter invariant results derived by taking a singular 
limit of de Sitter invariant gauge fixing functionals \cite{FH}. The
two methods are not completely discordant because both find the same
poles at $M_S^2 = 0$ in the spin two sector and (for $D= 4$) at 
$M_S^2 = -6/(2 - \beta) H^2$ in the spin zero sector. Moreover,
both methods give the same result for the linearized Weyl-Weyl
correlator \cite{Kouris}, after some mistakes are corrected in 
the invariant propagator computation \cite{Atsuchi}. So there does
not seem to be much support for the suggestion that the de Sitter
breaking propagators \cite{TW4,RPW,MTW3,KMW} access a different
sector of the graviton Hilbert space \cite{HMM}. 

It seems to us that the more likely source of the disagreement is 
the procedure of taking a singular limit of the invariant --- but 
provably wrong \cite{MTW1} --- propagators that derive from adding 
a de Sitter invariant gauge fixing functional to the action \cite{FH}. 
We conjecture that this introduces an error which only affects the 
de Sitter breaking, infrared divergent part of the propagator. This
error drops out of the linearized Weyl-Weyl correlator on account of
the four derivatives. (Note that even the provably wrong propagators
\cite{INVPROP} give the same result for the linearized Weyl-Weyl
correlator \cite{Kouris}.) One could check for such an error by
acting the gauge-fixed kinetic operator on the alleged propagator,
which should produce a projection operator. We predict that the
result will fail to be idempotent.

\vskip 1cm

\centerline{\bf Acknowledgements}

This work was partially supported by European Union program Thalis 
ESF/NSRF 2007-2013, by European Union Grant 
FP-7-REGPOT-2008-1-CreteHEPCosmo-228644, by NSF grants PHY-0855021 
and PHY-1205591, and by the Institute for Fundamental Theory at the 
University of Florida.

\end{document}